\documentclass[aps, prd, amsmath,amssymb,notitlepage, twocolumn, nofootinbib]{revtex4-1}

\usepackage{subfigure}
\usepackage{amssymb}
\usepackage{graphicx}
\usepackage{color}
\usepackage{amsmath, amsthm}
\usepackage[colorlinks=true,linkcolor=blue,citecolor=red,urlcolor=blue]{hyperref}
\usepackage{framed}

\newcommand{\be}{\begin{equation}}
\newcommand{\ee}{\end{equation}}
\def\mbf{\mathbf}
\def\tbf{\textbf}

\def\i{\text{i}}
\newcommand{\rt}{}

\begin{document}

\title{Spontaneous photon production in time-dependent epsilon-near-zero materials}

\author{A. Prain,  S. Vezzoli, N. Westerberg, T. Roger,  D. Faccio}
\affiliation{Institute of Photonics and Quantum Sciences, School of Engineering and Physical Sciences, Heriot-Watt University, EH14 7AS Edinburgh, UK.}

\begin{abstract}
	
 Quantum field theory predicts that a spatially homogeneous but temporally varying medium  will excite photon pairs out of the vacuum state. However, this important theoretical prediction lacks experimental verification due to the difficulty in attaining the required non-adiabatic and large amplitude changes in the medium.  Recent work has shown that in epsilon-near-zero (ENZ) materials it is possible to optically induce changes of the refractive index of the order of unity, in femtosecond timescales. By studying the quantum field theory of a spatially homogeneous, time-varying ENZ medium, we  theoretically predict photon pair production that is up to several orders of magnitude larger  than in non-ENZ time-varying materials. We also find that whilst in standard materials the emission spectrum depends on the time scale of the perturbation, in ENZ materials the emission is always peaked at the ENZ wavelength. These studies pave the way to technologically feasible observation of photon pair emission from a time-varying background with implications for quantum field theories beyond condensed matter systems and with potential applications as a new source of entangled light.

\end{abstract}

\date{\today}
\maketitle

{\bf{Introduction.}}  Epsilon-near-zero (ENZ) materials are characterised by  relative dielectric permittivity, whose real part, $\varepsilon_r$,  attains near-zero values around a given frequency $\omega_\textrm{ENZ}$ \cite{Engheta2013,opn}. A natural example of ENZ materials {that} are commercially available transparent conducting oxides, e.g. indium-tin-oxide (ITO) or Al-doped ZnO (AZO), where  $\varepsilon_{r}$ crosses zero near the plasma frequency, whereas the imaginary part $\varepsilon_{i}$  is small. These materials are attracting attention for their rather remarkable properties, ranging from a geometric invariance of resonant structures to novel light propagation regimes and light emission  geometries \cite{engheta1,opn,eps0_4,eps0_6,eps0_9,eps0_5,eps0_3,eps0_11,eps0_15}. Recent work pioneered by Engheta et al. has started to focus on the quantum properties of these materials, including quantum emission in ENZ cavities and limitation or even complete suppression of vacuum modes \cite{engheta2,engheta3}.\\
\indent Alongside the fascinating linear optical properties of these materials,  the nonlinear optical response exhibits an enhancement for frequencies close to  $\omega_\textrm{ENZ}$ \cite{Vincenti2011,Suchowski2013,Luk2015,Capretti2015,eps0_13,ciattoni:2010,Kinsey2015,eps0_13}. In these first studies, the main enhancement mechanism of the nonlinearity is related to an enhancement of the z-component (directed along the optical propagation axis) of the electrical field of the intense optical pump beam.\\
An alternative nonlinear enhancement mechanism has also been reported that is based on the simple realisation that in the ENZ region even a small change of permittivity $\Delta \varepsilon${,} induced by a third-order Kerr nonlinearity{,} can result in a large relative change of the refractive index $\Delta n/n_0${,} 
as a result of the small {\emph{linear}} refractive index $n_0$ \cite{Lucia,Boyd2016}. Indeed, for materials such as ITO and AZO that also exhibit a small imaginary part of the permittivity, $\varepsilon_i$, the real part of the refractive index $n_0$ is also close (although never equal) to zero around $\omega_\textrm{ENZ}$:  thus $\Delta n/n_0$ can easily attain values of the order of unity, compared to the modest $10^{-4}-10^{-3}$ values of non-ENZ materials \cite{Lucia,Boyd2016}. Remarkably these huge changes are also ultrafast, with rise-times below the duration ($\sim100$ fs) of the laser pulse used to induce the nonlinear response.\\
\indent The main underlying concept of this work is that the ultrafast and order-of-unity refractive index change in ENZ materials provides access to the quantum physics of spatially homogenous media with a time-varying parameter.\\
 \indent Quantum field evolution on a time-dependent background is a hallmark problem in quantum field theory (QFT) in curved spacetimes. 
 One of oldest and most robust predictions of QFT is that a homogeneously expanding universe will excite entangled photon pairs from the quantum vacuum state \cite{Parker_original, birrell1984quantum,Jacobson:2003vx, Ford:1997hb,mukhanov}. Although this kind of particle production is believed to be dominant in the very early universe and is argued to be responsible for the initial seeds of inhomogeneity{,} which gave rise to the visible structures of the universe we live in \cite{Mukhanov:1981xt,Martin2005, Martin2007}, no experimental evidence of this has been observed so far. A similar effect is predicted to occur also in condensed matter systems \cite{original_Visser} including Bose Einstein Condensates \cite{Fleshbach, BEC_cosmo,Prain:2010zq}, rings of trapped ions \cite{Ion_trap}, superfluid gases \cite{superfluid_gas} and non-linear optics \cite{Liberati:2011ep}. There is a specific and independent body of literature concerned with the classical and quantum consequences of a homogeneous time variation of the real part of the refractive index alone under the name `time refraction' \cite{Mendonca1,Mendonca2}.  Perhaps the first work in this general area is the work of Yablanovitch from 1989 \cite{yablanovitch}. \\
\indent In this work, we consider the specific case of the optical non-linear Kerr effect in an ENZ material, such that a passing pulse of intense light modifies the refractive index of the background material on top of which vacuum fluctuations are propagating.  
Using methods of QFT that have previously been applied in the context of expanding cosmologies, we explicitly calculate the photon pair production in ITO close to the ENZ frequency window. We show an  enhancement of seven orders of magnitude compared to non-ENZ Kerr media and demonstrate that the emission spectrum always shows a peak emission close to the ENZ wavelength (1377 nm in {our} studies). Taking losses into account by including the imaginary {part} of $n${,} {using the data of Ref.~\cite{boyd2003nonlinear}}
we predict roughly $10^{-4}$ photon pairs {per pump pulse} in 1x1x1 $\mu$m${}^3$ of ITO, which is in the detectable region of current single photon detectors.\\

\begin{figure}[t!] 
\centering 
\includegraphics[width=4.5cm]{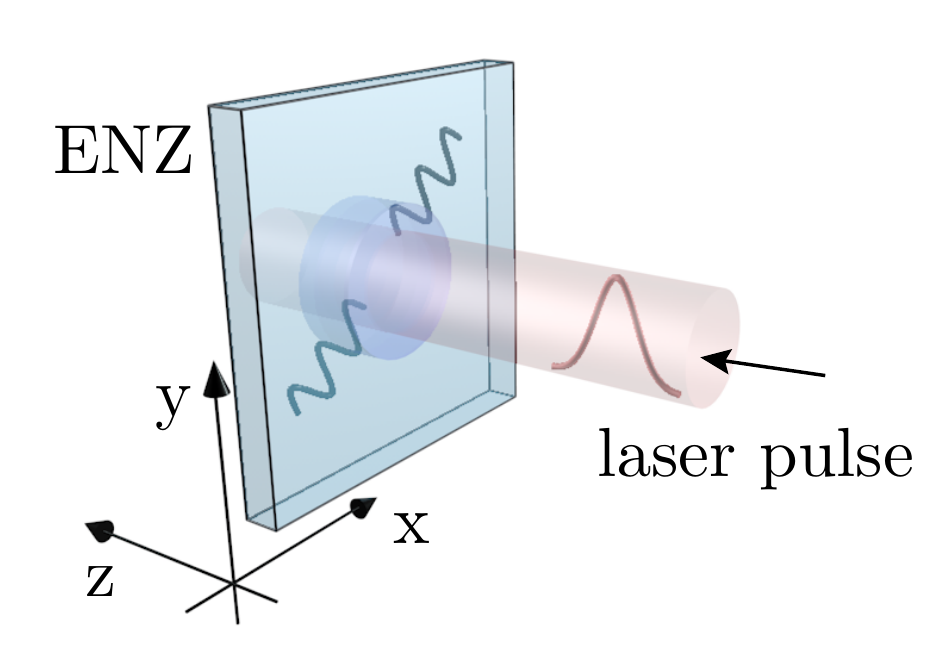}
\caption{Schematic layout of the interaction geometry: an intense laser pulse directed along the z-axis uniformly illuminates a region of the ENZ material. Photons are produced in the x-y plane in which the time-variation is spatially uniform. \label{F:layout}}
\end{figure}

{\bf{QFT model.}} We first illustrate how particle production emerges in a standard nonlinear medium as a consequence of a time variation of the dielectric environment induced by a pump beam.  In a spatially homogeneous material (i.e. pumped with a spatially uniform laser beam) that is changing in time, the $k$ vector is conserved and the optical frequency $\omega$ is modified. Spatial homogeneity allows one to separate the partial differential wave equation governing the full quantum field into a collection of ordinary differential equations (ODEs) labelled by $k$:
\be
\frac{\partial^2}{\partial t^2}E-c^2\nabla^2E\,\, \Rrightarrow \,\,  \frac{d^2 E_k}{dt^2}+\omega_k^2(t) E_k=0 \label{E:QHO}.
\ee
Eq.~\eqref{E:QHO} describes the evolution of a single fixed momentum mode $E_k(t)$ of a full quantum field $E(t,x)\propto\int \text{e}^{-\i k x}E_k(t)$ and here the time dependence of $\omega_k(t)=c k/n(t)$ is due to a time-dependent refractive index $n(t)=n_0+\Delta n(t)$, with $\Delta n(t)=n_2I(t)$, where $n_2$ is the nonlinear Kerr index and $I(t)$ is the intensity profile of the laser pump pulse \cite{boyd2003nonlinear}. We also assume that the medium is pumped orthogonally so that the time-variation is spatially uniform in the x-y plane: photons are then produced in this plane, as schematically summarised in Fig.~[\ref{F:layout}].

\begin{figure}
	\centering
	\includegraphics[width=8cm]{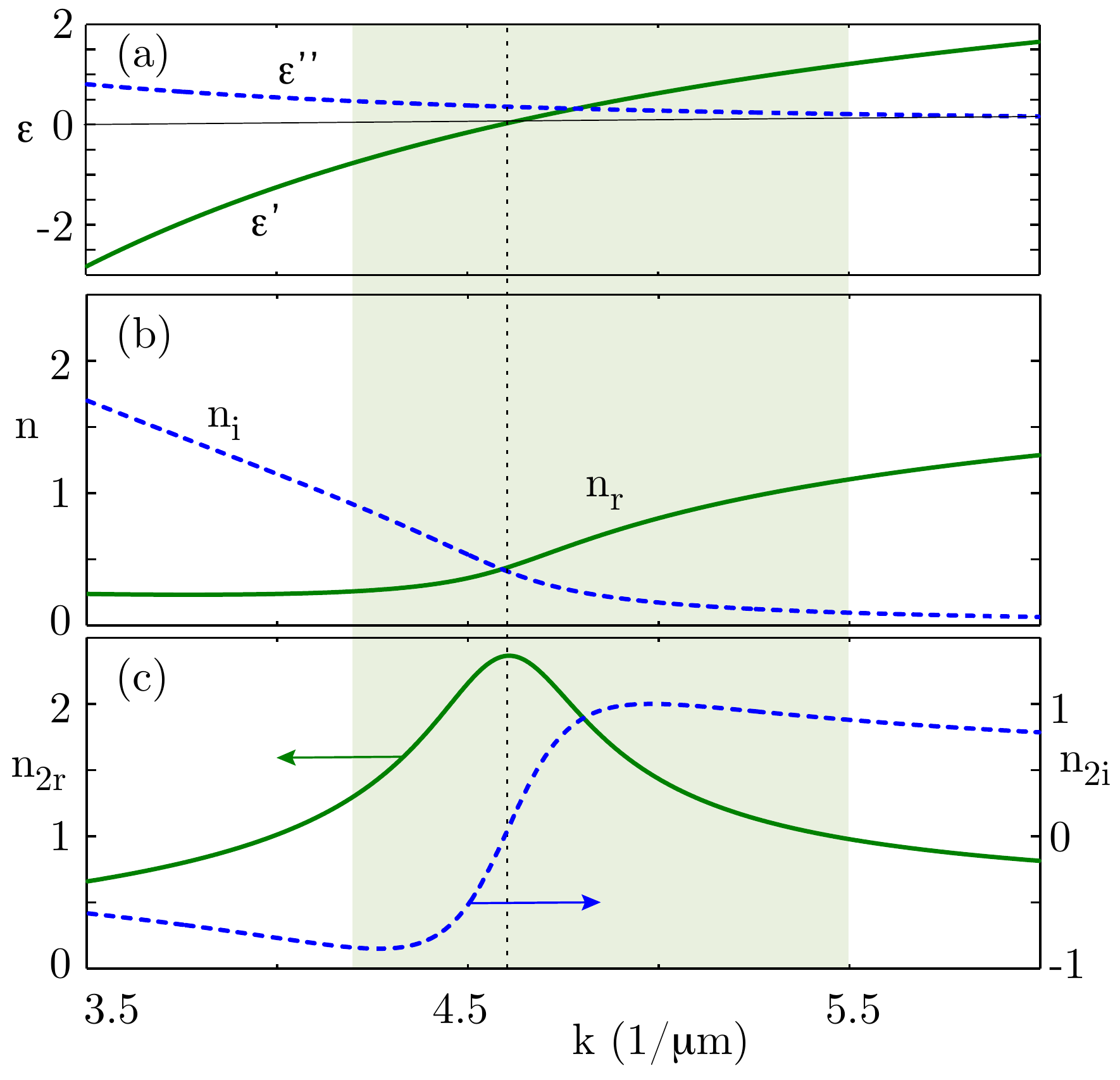}
	\caption{(a) Dielectric permittivity and (b) refractive index for ITO. (c)The nonlinear refractive indexes are represented by the proportionality factors $n_{2r}\propto{(n_r+n_i)}/{(n_r^2+n_i^2)}$ and $n_{2i}\propto{(n_r-n_i)}/{(n_r^2+n_i^2)}$. The shaded area is the ENZ region, here taken as as $|\varepsilon'|<1$.  The verical dotted line indicates the ENZ frequency.  \label{F:n}}
\end{figure}

The functions $\omega_k(t)$ are constant, $\omega_k=c k/n_0 $, well before ($t\rightarrow-\infty$) and after ($t\rightarrow\infty$)  the nonlinear interaction and change quickly in a small time window related to the pump pulse duration. The notion of particle or photon is only well defined in regions for which $\omega_k(t)$ is approximately constant $\dot{\omega_k}/\omega_k\ll1$ i.e. for modes in which the time variation is adiabatic, whereas in the intermediate regime, when $\omega_k(t)$ is varying rapidly, the notion of a photon is approximate or absent altogether \cite{Winitzki:2005rw}. 
Very generically speaking, an initial configuration of well defined photon number, when propagated through a (time) region in which the photon number is not defined can{,} and in general will{,} emerge into the second region where photon number is again well defined but with a different number of photons (see for example Ref.~\cite{birrell1984quantum} for an overview).\\ 
\indent For each momentum $k$ it is possible to calculate the total number of produced photons of momentum $k$ by solving the equation of motion Eq.~\eqref{E:QHO} with a purely positive frequency initial condition. This input condition evolves into a linear combination of positive and negative frequencies:
\be
\frac{\text{e}^{-\i\omega_k t}}{\sqrt{2\omega_k}}\stackrel{t\rightarrow -\infty}{ {\rt \xrightarrow{\makebox[1cm]{}} } }E_k(t)\stackrel{t\rightarrow +\infty}{ \xrightarrow{\makebox[1cm]{}} } \alpha_k\,\frac{\text{e}^{-\i\omega_k t}}{\sqrt{2\omega_k}}+\beta_k \frac{\text{e}^{+\i\omega_k t}}{\sqrt{2\omega_k}}.  \label{E:bogo}
\ee
The Bogoliubov coefficient $\beta_k$ is the coefficient of the negative frequency component of the solution at late times.
The number of produced photons for a given $k$ is then directly proportional to the square of the Bogoliubov coefficient $|\beta_k|^2$ \cite{birrell1984quantum}.\\
\begin{figure}
	\centering
	\includegraphics[width=8cm]{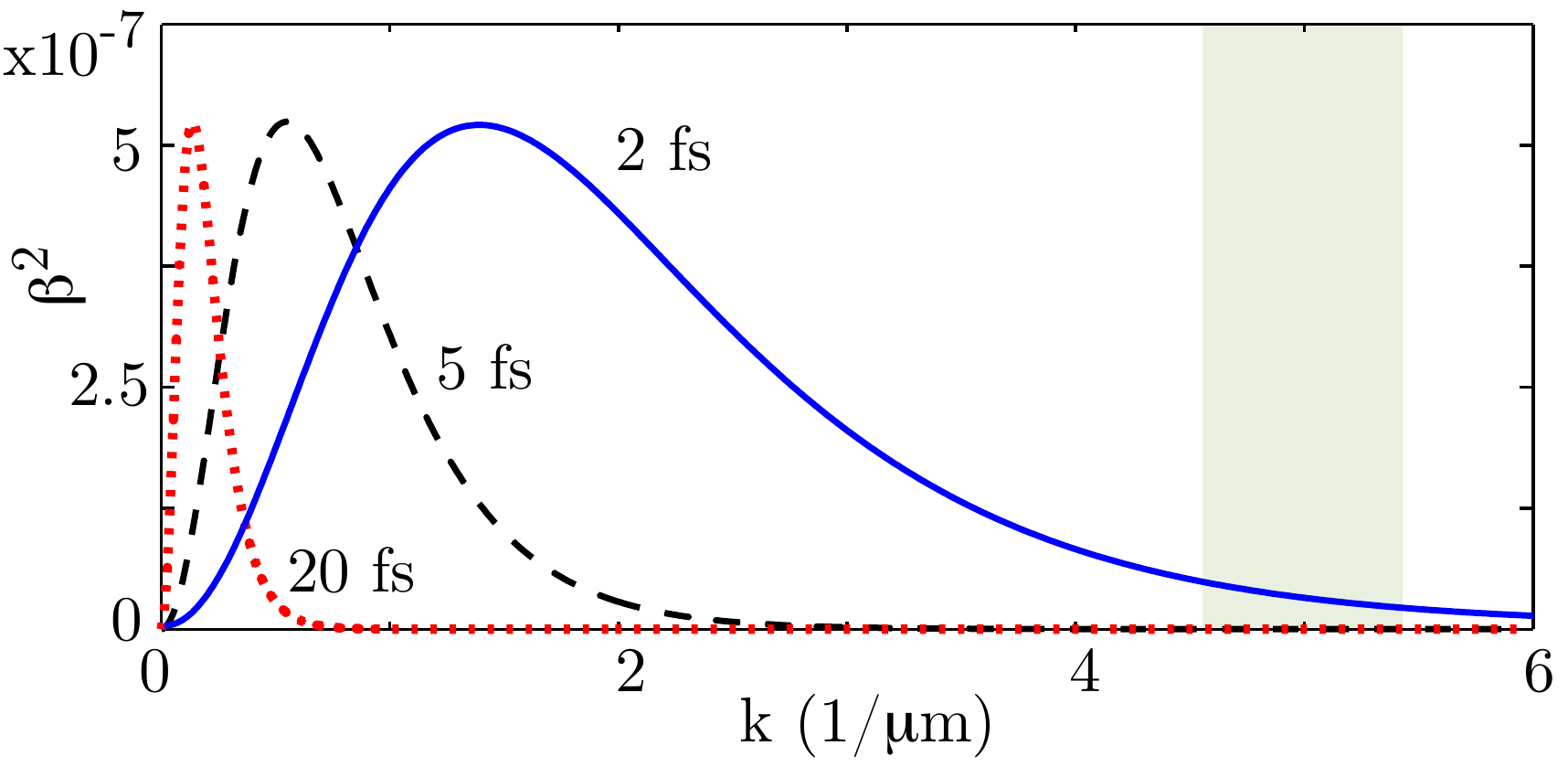}
	\caption{ The spectrum of produced photons ($\propto \beta^2$) in the dispersion-less case ($n_0=1$), the material response is assumed constant in $k$ and such that $\Delta n/n_0=10^{-3}$. Our equations reproduce the standard result showing how the peak of the spectrum shifts to the infrared for increasing time-variation rise time $\tau$, indicated in the figure.  \label{F:no_dispersion_combined}}
\end{figure}
\indent For illustration, consider Eq.~\eqref{E:QHO} with a time dependent frequency given by $\omega_k(t)=ck/n(t)$ with refractive index $n(t)=1+\delta \,\text{sech}^2({t}/{\tau})$ where $\delta=\Delta n/n_0$.  In this case the Bogoliubov spectrum admits an analytical solution (see supplemental material) and takes the shape shown in Fig.~[\ref{F:no_dispersion_combined}]: the emitted photon spectrum has a peak that shifts to the infrared for increasing rise-time $\tau$ of the time-variation, where single or few photon detection is extremely challenging or currently not possible. In the experimentally feasible detection region (for example in the green shaded region, taken as the same ENZ region examined in more detail below) the amplitude $|\beta_k|^2$ is exponentially suppressed. Finally, the maximum number of photons is typically extremely small also because the maximum amplitude of $|\beta_k|^2$ scales with the (squared) variation of the medium, $\delta^2$ (see supplementary material), which is very small ($\delta^2\simeq 10^{-6}$) in typical nonlinear media. All combined, these known results lie at the heart of the problem of actually detecting photon emission from a standard time dependent medium. \\
\noindent {\bf{Photon-pair creation in the ENZ region.}}  
{In} an ENZ material, close to the ENZ frequency it is possible to use a simple resonance-free Drude-Lorentz model for the linear permittivity \cite{Luk2015,Boyd2016}:
\be
\varepsilon'+\i\varepsilon''=\varepsilon_\infty-\frac{\omega_p^2}{\omega^2+\i\omega\Gamma} ,\label{E:drude}
\ee
where $\varepsilon'$ and $\varepsilon''$ are the real and imaginary parts of the permittivity, $\Gamma$ is related to the electron damping in the material and $\omega_p$ is the plasma frequency. Essential in this work is the existence of a zero crossing for the real part of the permittivity, which occurs at approximately $\omega_\text{ENZ}\simeq \sqrt{\omega_p^2-\Gamma^2}$. In Ref.~\cite{Luk2015}, experimental values for a ITO are reported as $\varepsilon_\infty\simeq 4.082$, $\omega_p^2\simeq 7.643\times 10^{30}$s${}^{-1}$ and $\Gamma\simeq1.239\times 10^{14}$s${}^{-1}$ leading to $\omega_\text{ENZ}$ at around 1.36$\times 10^{15}$s${}^{-1}$ or, in terms of wavelength, at 1377 nm. Near the ENZ region, the real part of the refractive index defined through $n_0=n_{0r}+\i n_{0i}=\sqrt{\varepsilon'+\i\varepsilon''}$ is also close to zero (but never exactly zero due to the non-zero imaginary permittivity), as shown in Fig.~[\ref{F:n}].

In a dielectric dispersive medium, it is possible to reduce the full the wave equation to the study of equations of the form \eqref{E:QHO} (see supplementary information):
\be
\frac{\partial^2}{\partial t^2} E_k+\omega_k^2 E_k=0 , \qquad  \omega^2_k=\frac{c^2K^2}{\varepsilon'+\i\varepsilon''} \label{E:start}
\ee
where $K$ is the material wavenumber{, which is} 
conserved during the time evolution{. Expressed} in terms of the vacuum wavenumber{, the material wavenumber} is $K(k)=k\left[n_{0r}(ck)+\i n_{0i}(ck)\right]$, where $n_{0r}$ and $n_{0i}$ are the constant, real and imaginary parts of the background refractive index. Hence, the relevant dynamical equation is of the form Eq.~\eqref{E:QHO} with 
\be
\omega_k(t)=\frac{cK(k)}{n_{r}(ck)+\i n_{i}(ck)} \label{E:freq}
\ee
 The time dependence in the frequency of Eq.~\eqref{E:freq} is  given by the nonlinear change of the refractive indices which appear in the denominator. 
 In particular the real part of the refractive index $n_r$ is modified due to the pump laser pulse via the nonlinear Kerr effect so that $n_r(t)=n_{0r}+n_{2r}I(t)$ where $I(t)$ is the local intensity of the pump beam and $n_{2r}$ is the real part of the non-linear Kerr index. 
We recall that in ENZ materials the nonlinear response is strongly affected by the linear dispersion \cite{Lucia}. As a consequence, the nonlinear change of the real part, $\Delta n_r=n_{2r}I(t)$ can be of the order of unity{.}
{Assuming} that $\chi^{(3)}$ has no dispersion{, and has} equal real and imaginary parts, $n_{2r}\propto D=(n_{0r}+n_{0i})/(n_{0r}^2+n_{0i}^2)$ which is peaked around the ENZ frequency \cite{Lucia} (see Fig.~[\ref{F:n}]).\\
\begin{figure}[t!]
\centering
\includegraphics[width=8cm]{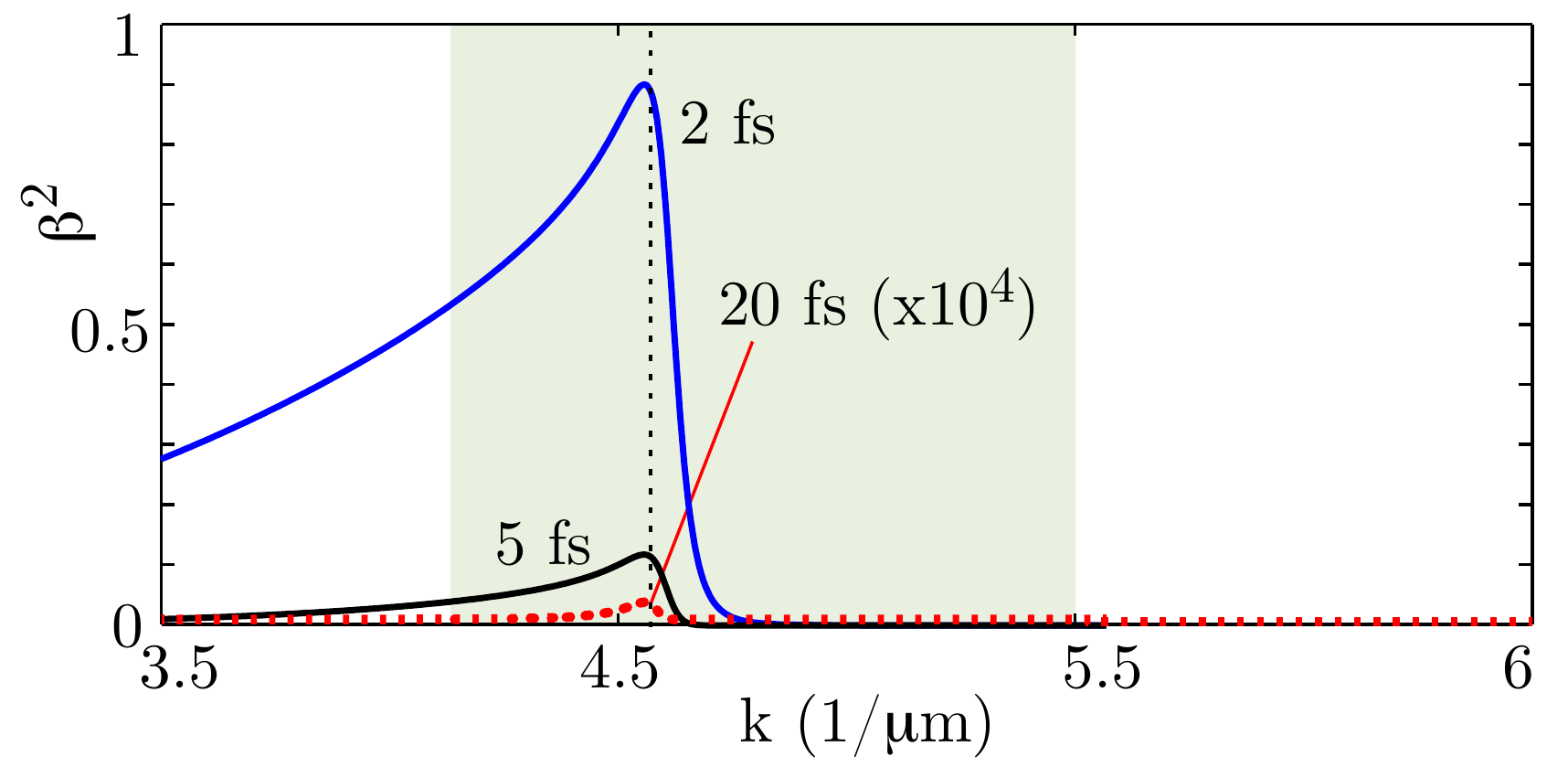}
\caption{ The spectrum of produced photons in an ENZ material from the full calculation assuming that the imaginary part of the refractive index is zero. The shaded region coincides with the ENZ region  region of Fig.~[\ref{F:n}]. Different rise times are indicated in the figure. The vertical dotted line indicates the ENZ frequency.  \label{F:ni=0}}
\end{figure}
{{\indent The frequency given by Eq.~\eqref{E:freq} is real valued for early and late times where it reduces to $\omega_k=ck$. However, during the time-variation, $I(t)\neq0$ and therefore $\omega_k(t)$ becomes complex. {The refractive indices} $n_r$ and $n_i$ both enter in the real and imaginary parts of $\omega_k(t)$; 
including $n_i$ leads to a modification of the particle production spectrum (governed by the real part of $\omega_k(t)$) and to a time modulated absorption (governed by the imaginary part of $\omega_k(t)$). We note that $n_{2i}\propto(n_{0r}-n_{0i})/(n_{0r}^2+n_{0i}^2)$  \cite{Lucia}, which therefore is zero exactly at the ENZ frequency ($n_{0r}=n_{0i}$, see Fig.~[\ref{F:n}(c)]).  Its effect will therefore be felt inside the ENZ region but not at the exact ENZ frequency.
\\ }
In the following, we model the laser pump pulse and hence the temporal variation profile with a sech${}^2(t/\tau)$ function and we numerically calculate the spectrum for various rise times $\tau$ and with a positive frequency input condition as indicated in Eq.~\eqref{E:bogo}. \\
{\bf{Results.}} As a first calculation{,} let us assume that $n_i=0$. The numerical results are shown in Fig.~[\ref{F:ni=0}]. Comparing to Fig.~[\ref{F:no_dispersion_combined}], we note a first and obvious difference: although in both figures we use the same values for the time-variation rise times ($\tau=$ 2, 5 20 fs), in the case of an ENZ material the typical and expected red-shift of the photon emission peak for increasing $\tau$ is completely absent. Rather, we find a new emission peak that is now locked onto the ENZ frequency. This is somehow reminiscent of a similar frequency locking observed for resonant antennas placed in proximity to an ENZ substrate \cite{antenna,antenna2} and is of crucial importance for future experiments: ENZ allows one to separate the temporal rise time of the pump laser pulse from the peak spectral emission wavelength. In this way, longer and technologically accessible laser pulses can be used in
combination with the possibility to match the emission to the photon detection sensitivity (typically centred around 700 nm or 1400 nm wavelengths). We also underline the significant enhancement in the emissivity that is now close to seven orders of magnitude better if we compare the absolute maxima in both cases but is actually many orders of magnitude larger if we restrict the comparison to the ENZ window{,} and for pump laser pulses with technologically feasible durations of 5 fs or larger. \\
\begin{figure}[t!]
\centering
\includegraphics[width=8cm]{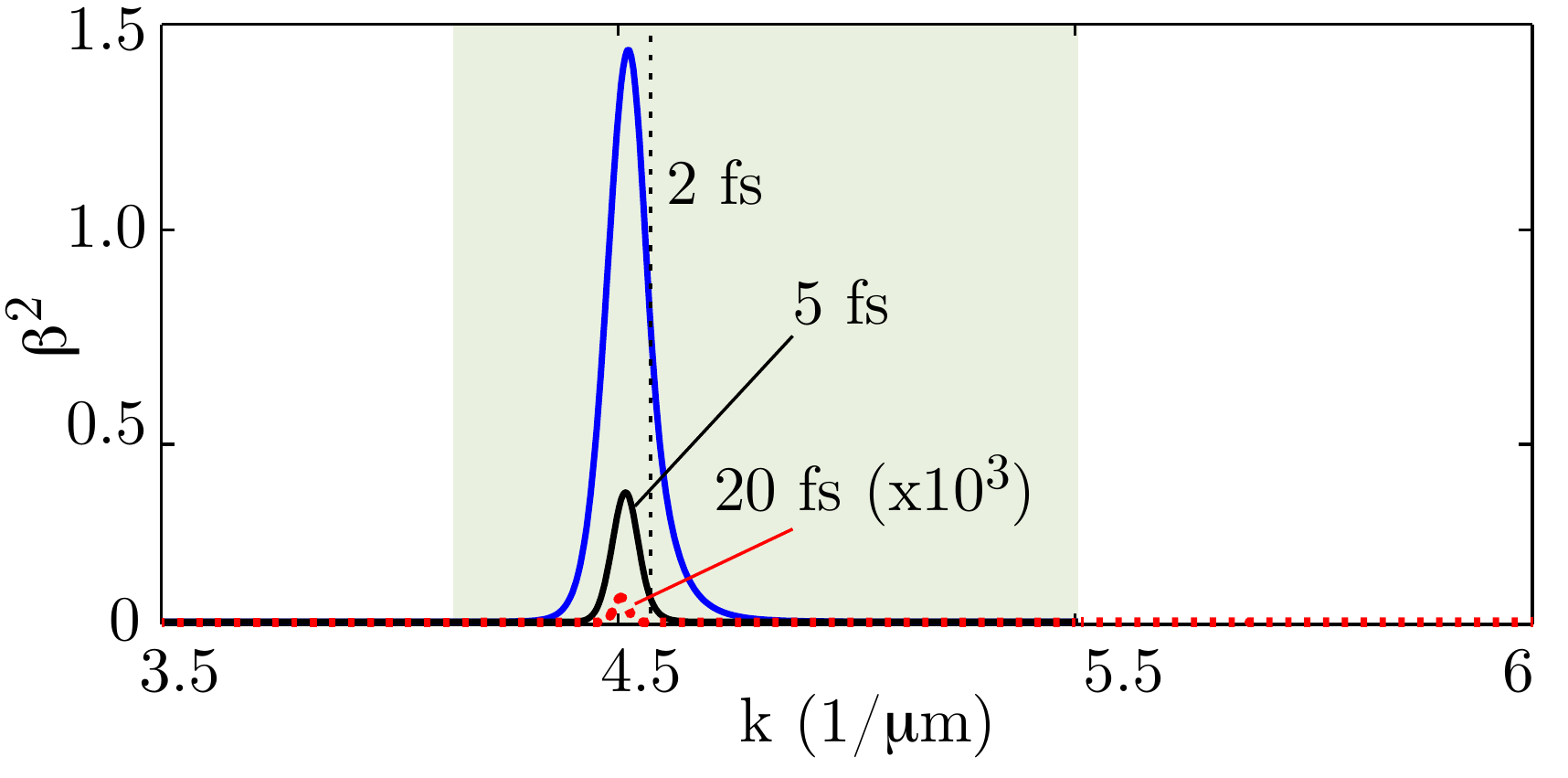}
\caption{The spectrum of produced photons in an ENZ material from the calculation with the full complex refractive index shown in Fig. 3b). Different rise times are indicated in the figure. \label{F:all_in}}
\end{figure}
\indent {{In Fig.~[\ref{F:all_in}] we show the results using the full complex refractive index, with real and imaginary parts shown in Fig.~\ref{F:n}(b). The maximum values of $|\beta_k|^2$ are sightly enhanced by the additional contribution of $n_i$. The peak is now also relatively localised around ENZ due to the fact that the real part of $\omega_k$ that is responsible for photon production, depends on both $n_r+n_{2r}(t)$ and $n_i+n_{2i}(t)$ (see supplementary material). Therefore, for frequencies above ENZ, the time variation of $\omega_k$ occurs on a high $n_r$ background. For low frequencies, the time variation occurs on a high $n_i$ background (see Fig.~[\ref{F:n}]). Only very close to ENZ, the time variation is enhanced by a simultaneously low $n_r$ and $n_i$. }}
A smaller effect is also visible: the peak emission is slightly shifted to lower frequencies with respect to the ENZ frequency (vertical dotted line). This is a result of  the $n_i$ time-varying contribution being equal to zero exactly at the ENZ frequency and asymmetric around ENZ.. \\
\indent In terms of the number of photons actually emitted, we must account also for propagation absorption of these photons{. This} 
leads to an additional exponential damping term that restricts the allowed physical size of the medium. For example, if
we consider, a 1x1x1 $\mu$m$^3$ volume, then the effects of losses will amount to a $\sim10$x reduction factor with respect to the lossless case.
We therefore predict an actual photon emission spectrum of $N_k\sim10^{-6}|\beta_k|^2$ per pump laser pulse (see supplementary material): a 5 fs (10 fs when measured at full width) laser pulse is expected to produce $\sim10^{-7}$ photon pairs at the ENZ wavelength per laser pulse, which would imply a $\sim10$ Hz emission rate using a 100 MHz laser source. These photons could then be extracted from the thin film, either by a grating coupler or by direct coupling to a waveguide structure.\\
\noindent {\bf{Conclusions.}}
ENZ materials exhibit a series of novel nonlinear features that lead to a dramatically enhanced interaction with the quantum vacuum. 
We demonstrate a seven-orders of magnitude enhancement of the quantum amplification of vacuum fluctuations for (vacuum) frequencies close to the ENZ region, i.e. in the near IR. We describe a scenario where the detection of pairs of entangled photons should be experimentally feasible. 

The vacuum-seeded photon emission due to a time-varying ENZ medium presented here is very generic. It is independent on the specific choice of ENZ material and functional shape of the time variation: these parameters may be tuned to increase the expected photon emission rate although we preferred to choose physically relevant parameters, i.e. as measured for a real material. 
The emission at the ENZ wavelength should be easily distinguishable from other effects such as self-phase-modulation on the pump pulse (by pumping at a distant wavelength) but fluorescence emission will need to be assessed to ensure that it is negligible in the ENZ region. 
 We also note that, in keeping with the standard quantum optics description \cite{quantum,quantum2}, a non-zero $\beta_k$ Bogoliubov coefficient implies optical amplification, i.e. a (classical laser pulse) seeded amplification process could be used to assess the effects of the time-varying ENZ material.\\

{\bf{Acknowledgements.}} A.P. acknowledges financial support the European Union's Horizon 2020 research and innovation programme under grant agreement No 659301 and S. Seahra for help with the numerical code. D. F. acknowledges financial support from the European Research Council under the European Union Seventh Framework Programme (FP/2007-2013)/ERC GA 306559 and EPSRC (UK, Grant No. EP/M009122/1). {\rt N.W. acknowledges support from the EPSRC CM-CDT Grant No. EP/L015110/1.}


%

\newpage
\section{Supplementary Material}

\noindent {\bf{Explicit functional dependence of $|\beta_k|^2$ for a non-dispersive medium.}} \\
  A simple model for the pump induced time dependence makes use of a time dependent refractive index as  
\be
\partial_t^2 E_k+\frac{c^2k^2}{\left(n_0+\delta n(t)\right)^2}E_k=0.
\ee
where $n_0$ is assumed to be a constant and $\delta n$ is only a function of time.  Exact results \cite{Boonserm:2010px} for the emission spectrum are available for special choices of the perturbation $\delta n(t)$. For example for the choice $\delta n(t)=\delta\,\text{sech}^2(t/\tau)$ with $\delta\ll1$ the exact result is
\begin{align}
|\beta_k|^2&=\frac{\cos^2\left(\frac{\pi}{2}\sqrt{1-\frac{8c^2k^2\delta\tau^2}{n_0^3}}\right)}{\text{sinh}^2\left(\frac{\pi ck\tau} {n_0}\right)} \notag \\
&\simeq \frac{4\pi^2c^4k^4\tau^4}{n_0^6\,\text{sinh}^2\left(\frac{\pi ck\tau}{n_0}\right) }\, \delta ^2+\mathcal{O}\left(\delta^3\right) \label{E:result}.
\end{align}
The peak of the spectrum occurs at  $k_\text{peak}\simeq (c\tau)^{-1}$ as we anticipated and, inserting this into the result \eqref{E:result} and choosing $n_0\simeq 1$, we find $|\beta_{k_\text{peak}}|^2\simeq 0.7\times \delta^2$ as we stated in the body of the work.

\vspace{2mm}

\noindent \tbf{Derivation of Eq.~\eqref{E:start} as an effective equation of motion}\\
 Starting {from} the macroscopic Maxwell equations{,} it is {straightforward} to derive the integro-differential equation satisfied by the electric field
\be
-\nabla^2 E+\frac{1}{c^2}\partial_t^2\left(E(t,x)+\int^t_{-\infty} \chi(t-s)E(s,x)ds\right)=0. \label{E:wave1}
\ee
where $\chi$ is the susceptibility. The non-locality in time represented by the convolution integral
is responsible for all dispersive effects. { Physically, it is due to the delayed response of the charges inside a dielectric to the propagating electric field}. A well-known {and} simple model for this dispersion is given by the so-called Drude-Lorenz model{,} which can be expressed in frequency space as
\be
\chi(\omega)=\left({\varepsilon_\infty}-1\right)-\frac{\omega_p^2}{\omega^2+\i\Gamma\omega}. \label{E:suscep}
\ee
{ Here} $\omega_p$ is the plasma frequency, $\Gamma$ is a coefficient representing the damping in the motion of the charges inside the material { and $\varepsilon_\infty$ is the permittivity at infinite frequency. The} 
equation of motion \eqref{E:wave1} { then} becomes
 \be
 \nabla^2E_\omega+\frac{\omega^2}{c^2}\left(\varepsilon_\infty-\frac{\omega_p^2}{\omega^2+\i\Gamma\omega}\right)E_\omega = 0.
 \ee
Taking a further, spatial Fourier transform one arrives at the dispersion relation satisfied by pairs $(\omega,K)$ representing propagating degrees of freedom
\be
-K^2+\frac{\omega^2}{c^2}\left(\varepsilon_\infty-\frac{\omega_p^2}{\omega^2+\i\Gamma\omega}\right)=0 \label{E:dis}.
\ee
Across a material boundary the frequency is conserved but the wave number is replaced by the complex quantity
\be
k\longrightarrow K(\omega)=\frac{\omega}{c}\left[n_r(\omega)+\i n_i(\omega)\right],
\ee
which accounts for refraction across the spatial boundary and the onset of absorption under further propagation inside the material.  This relation is obtained by solving \eqref{E:dis} for $K$ as a function of $\omega$. Given that the frequency is conserved across the material boundary{,} we can substitute its vacuum value $\omega=ck$ into { the above }
relation{. This leads} 
to the relationship between in-material and vacuum wavenumber
\be
k\longrightarrow K(ck)={k}\left[n_r(ck)+\i n_i(ck)\right]\label{E:refrac}.
\ee
{In order to capture that the frequency is conserved across a material boundary, we can now perform the}
same substitution $\omega=ck$ into the susceptibility \eqref{E:suscep} 
\be
\chi(k)=\varepsilon_\infty-\frac{\omega_p^2}{c^2k^2+\i\Gamma ck}.
\ee
{This simplifies Eq.~\eqref{E:dis}}
to 
\be \label{E:dis1}
-\omega^2 +\frac{c^2K^2}{\left(\varepsilon_\infty-\frac{\omega_p^2}{c^2k^2+\i\Gamma ck}\right)}=0.
\ee
{Note that the above can be simplified further to give the expected $\omega = ck$. Let us however keep this in the form of Eq.~\eqref{E:dis1} for reasons given below. Now, to connect this to the main text, we first rewrite Eq.~\eqref{E:dis1} in}
the notation of the main text
\be
-\omega^2+\frac{c^2 K^2}{\varepsilon'(ck)+\varepsilon''(ck)}=0,
\ee
or in terms of the refractive index
\be
-\omega^2+\frac{c^2 K^2}{\left(n_r(ck)+\i n_i(ck)\right)^2}=0.
\ee
We will drop the explicit reference to the vacuum momentum $k$ in $n_r(ck)$, $n_i(ck)$ and simply write $n_r$ and $n_i$ in what follows. One may re-write this last equation as a 2nd order ODE in time for an electric field mode of vacuum momentum $k$
\be
\partial_t^2E_k+\frac{c^2K^2}{\left(n_r+\i n_i\right)^2}E_k=0 \label{E:linear}
\ee
At this stage we have employed no approximations: this differential equation is exactly satisfied by a plane wave of the form $E=\text{exp}(-\i\omega t+\i K x)$ inside the material.  We { will now} assume that the effective equation of motion for the electric field mode $E_k$ { in a (non-linearly-induced) time-varying medium is given by simply replacing the refractive indices in Eq.~\eqref{E:linear} by their non-linear time dependent ones:}
\be
\partial_t^2E_k+\frac{c^2K^2}{\left(n_r(t)+\i n_i(t)\right)^2}E_k=0,
\ee
where the time dependent refractive indices are given by the linear plus the time dependent non-linear correction
\be
n_r(t)=n_r+n_{2r}(t), \quad n_i(t)=n_r+n_{2i}(t).
\ee
Thus we arrive at Eq.~\eqref{E:start} in the text:
\be
\frac{\partial^2}{\partial t^2} E_k+\omega_k^2 E_k=0 , \qquad  \omega_k(t)=\frac{cK}{n_{r}+\i n_{i}} .
\ee
We note that the real part of $\omega_k$  is given by
\be
ck\frac{\left(n_r+n_{2r}(t)\right)n_r+(n_i+n_{2i}(t))n_i}{(n_r+n_{2r}(t))^2+(n_i+n_{2i}(t))^2}
\ee
It is the time dependence of this component that gives rise to photon production.

\vspace{2mm}


\noindent{\bf{From $\beta_k$ to photon number $N_k$.}}  \\
 In the body of this work we have outlined the scattering problem that needs to be solved in order to find the coefficient $\beta_k$. In order to go from this to the actual particle number spectrum three things need to be taken into account: The phase space volume, the physical size of the emitting region and the mode damping. 

In general, the Bogoliubov coefficient $\beta_k$ is related to the total number $N$ in D+1 dimensions by
\be
\frac{dN}{d\text{Vol}}=\int |\beta_k|^2d^D\mbf{k}
\ee
The parameter $k$ actually labels a large number of momenta all with the same magnitude but different direction. In D space dimensions the raw spectrum given by $|\beta_k|^2$ needs to be multiplied by the surface area of the D-1 sphere in k-space which is always equal to $k^{D-1}(2\pi)^{D/2}/\Gamma(D/2)$. In our case we are in $D=2$ so this is simply $2\pi k$. Then we have
\be
\frac{dN}{d\text{Vol}}=\int 2\pi k \,|\beta_k|^2d^2k
\ee

For the partial volume factor we recall that $|\beta_k|^2$ is actually measuring a particle number \textit{density} so that we need to multiply by physical volume to get the total particle number. We take the orthogonal dimension of the sample to be on the order of $L=1\mu$m so that the orthogonal volume is given by Vol$=\pi L^2=3\times 10^{-12}$m${}^{2}$.

In order to account for mode damping inside the material we have assumed that the imaginary part of the refractive index is constant.  To obtain an order of magnitude estimate of the effect of the damping factor we note that the modes decay approximately according to $\text{exp}(-\omega n_i x/c)$.  A typical value for $n_i$ for the ENZ material studied here is $n_i\simeq10^{0}$ to $10^{-1}$ so that this exponential factor is about $\text{e}^{-1}$ which contributes a factor $\text{e}^{-2}\sim10^{-1}$ to the Number${}_k$ values{,} since it's the square $|\beta_k|^2$ which counts photons.

The combined effect of these factors is given by
\begin{align}
N_k&=2\pi k\times \pi L^2\times 10^{-1}|\beta_k|^2 \notag \\
&=2\pi^2\times \left(4\times 10^6\right)\times \left(10^{-6}\right)^2\times 10^{-1} |\beta_k|^2\notag \\
&\simeq 10^{-6}|\beta_k|^2 \label{E:number_k}
\end{align}

This gives the conversion between $\beta^2$ and the actual {\emph{emitted}} photon number near the ENZ frequency, $k=4\times 10^6$m${}^{-1}$.

\end{document}